\input harvmac.tex
\input epsf





\line{\hfill PUPT-1763}
\vskip 1cm

\Title{}{Duality of large $N$ Yang-Mills Theory on $T^2 \times R^n$.}

\centerline{$\quad$ { Z. Guralnik}}
\smallskip
\centerline{{\sl Joseph Henry Laboratories}}
\centerline{{\sl Princeton University}}
\centerline{{\sl Princeton, NJ 08544, U.S.A.}}
\centerline{{\tt zack@puhep1.princeton.edu}}

\vskip .3in 

We find aspects of electrically confining large $N$ Yang-Mills 
theories 
on $T^2 \times R^{d-2}$ which are consistent with a $GL(2,Z)$ duality.
The modular parameter associated with this $GL(2,Z)$
is given by ${m\over N} + i\Lambda^2 A$, where $A$ is the area of the
torus, 
$m$ is the t'Hooft twist on the torus, and $\Lambda^2$ is the string 
tension. 
$N$ is taken to infinity 
keeping $m\over N$ and $g^2N$ fixed. 
This duality 
may be interpreted as T-duality of the QCD string if one identifies
the magnetic flux with a two-form background in the string theory.
Our arguments make no use of supersymmetry.
While we are not able to show that this is an exact self duality of 
conventional QCD,  we conjecture that it may be applicable within
the universality class of QCD. 
We discuss the status of the conjecture for
the soluble case of pure two dimensional Euclidean QCD on $T^2$,  
which is almost but not exactly self dual.  
For higher dimensional theories,  we discuss qualitative features
consistent with duality.
For $m=0$,  such a duality would lead to an equivalence
between pure QCD on $R^4$ and QCD on $R^2$ with two adjoint 
scalars.   When $\Lambda^2 A << m^2/N^2$,
the proposed duality includes exchanges of rank with
twist.  This exchange bears some resemblance,  but is not equivalent, 
to Nahm duality.  A proposal for an explicit perturbative map which 
implements duality in this limit is discussed.



\Date{}

\lref\eguchi{T. Eguchi and H. Kawai, {\it ``Reduction of dynamical 
degrees
		of freedom in the large N gauge theory,''} 
		Phys. Rev. Lett.48 (1982) 1063.}
\lref\nahm{ P. J. Braam and P. van Baal, {\it ``Nahms transformation 
	    for Instantons,''} Commun. Math. Phys. 122 (1989) 267; 
	   P. van Baal, {\it ``Instanton moduli for $T^3 \times R$,''}
	   Nucl. Phys. Proc. Suppl. 49 (1996) 238, hep-th/9512223.}
\lref\verlindetduality{H. Verlinde, talk given at Princeton 1996.}
\lref\dougmoore{M. Douglas and G. Moore, {\it ``D-branes, quivers and 
                ALE instantons,''}hep-th/9603167.}
\lref\EWbstate{E. Witten, {\it ``Bound states of strings and 
p-branes,''}
		Nucl. Phys. B460 (1996) 335, hep-th/9510135.}
\lref\polchbranes{ }
\lref\raminst{ M. Douglas, {\it "Branes within branes,"}
               hep-th/9512077.}
\lref\DBI{ R. G. Leigh, {\it "Dirac-Born-Infeld action from 
             Dirichlet sigma model,"} 
           Mod. Phys. Lett.A4 (1989) 2767.}
\lref\nonlinsusy{ J. Harvey and G. Moore, {\it "On the algebras
		   of BPS states,"} hep-th/9609017}
\lref\verlindehac{ H. Verlinde and F. Hacquebord, 
		  {\it "Duality symmetry of $N=4$ Yang-Mills
			theory on $T^3$,"} hep-th/9707179.}
\lref\zbranes{ Z. Guralnik and S. Ramgoolam, { \it "From
		0-Branes to Torons,"} hep-th/9708089.}
\lref\bfss{ T. Banks, W. Fischler, S. Shenker, L. Susskind,
		{\it "M Theory as a matrix model: a conjecture,"}
		 Phys. Rev. D55 (1997) 5112-5128, hep-th/9610043.}
\lref\lcone{ L. Susskind, {\it "Another conjecture about M(atrix)
		theory,"} hep-th/9704080.}
\lref\blau{ M. Blau and M. O'Loughlin, {\it "Aspects of U-duality
		in Matrix Theory,"} hep-th/9712047.}
\lref\hull{ C. Hull, {\it "U-duality and BPS spectrum of Super 
		Yang-Mills theory and M theory,"} hep-th/9712075.}
\lref\joao{ Private communication }
\lref\rabinov{ N. A. Obers, B. Pioline, E. Rabibovici, {\it 
		"M theory and U-duality on $T^d$ with gauge 
		backgrounds,"} hep-th/9712084.}
\lref\noncom{ A. Connes, M. Douglas, A. Schwarz, 
	{\it "Noncommutative geometry and Matrix theory:
	 compactification on tori,"} hep-th/9711162.}
\lref\hulldoug{ M. Douglas and C. Hull, {\it "D-branes and
		the noncommutative torus,"} hep-th/9711165.}
\lref\thooft{ G. 't Hooft, {\it "A property of electric and
		magnetic flux in non-abelian gauge theories,"}
	       Nucl. Phys. B153 (1979) 141-160.}
\lref\rudd{ R. Rudd, {\it "The string partition function
		for QCD on the torus,"} hep-th/9407176.}
\lref\gtaylor{ D. Gross and W. Taylor, {\it 
		"Two dimensional QCD is a string
		theory,"} Nucl. Phys. B400 (1993) 181,
		hep-th/9301068.}
\lref\gross{ D. Gross, {\it "Two dimensional QCD as a string
		theory,"} Nucl. Phys. B400 (1993) 161,
		hep-th/9212149.}
\lref\polyakov{A. Polyakov, {\it ``String representations and hidden
		symmetries for gauge fields,''} 
		Phys. Lett. 82B (1979) 247-250.
                {\it ``Gauge fields as rings of glue,''}
		Nucl. Phys.B164 (1980) 171-188.} 
\lref\igor{K. Demeterfi, I. Klebanov,  Gyan Bhanot, 
	   {\it ``Glueball spectrum in a $(1+1)$ dimensional 
	     model for QCD,''} Nucl. Phys.B418 (1994) 15-29, 
hep-th/9311015.}
\lref\daleykleb{S. Dalley, I. Klebanov, {\it String spectrum of 
$(1+1)$
	        dimensional large $N$ QCD with adjoint matter,''}
	        Phys. Rev. D47 (1993) 2517-2527.}
\lref\antonuccio{F. Antonuccio and S. Dalley, {\it `` Glueballs
		 from (1+1) dimensional gauge theories with 
		transverse degrees of freedom,''} 
                 Nucl. Phys. B461 (1996) 275-304, hep-ph/9506456.}
\lref\migdal{A. Migdal, {\it ``Recursion equations in gauge 
theories,''}
             Sov.Phys.JETP42 (1975) 413, Zh.Eksp.Teor.Fiz.69 (1975) 
810-822.}
\lref\witten{E. Witten, {\it ``Two dimensional gauge theories 
revisited.''}
	      J. Geom. Phys. 9 (1992) 303-368, hep-th/9204083.}
\lref\taylor{W. Taylor, IV, {\it ``D-brane field theory on 
		compact spaces,''} Phys. Lett. B394 (1997) 283-287,
		hep-th/9611042.}
\lref\sav{S. Sethi and L. Susskind, {\it ``Rotational invariance in 
	   the M(atrix) formulation of IIB theory,''} 
          Phys. Lett. B400 (1997) 265-268, hep-th/9702101.}
\lref\torons{Z. Guralnik and S. Ramgoolam, 
		{\it ``Torons and D-brane bound states,''}
	     Nucl. Phys. B499 (1997) 241-252, hep-th/9702099.}
\lref\tdlty{A. Giveon, M. Porrati, and E. Rabinovici, 
	{\it ``Target space duality in string theory,''}
 	 Phys. Rept. 244 (1994) 77-202, hep-th/9401139.} 
\lref\tseytlin{   }
\lref\blautom{M. Blau and G. Thompson, {\it ``Lectures on 2-D gauge 
	theories: topological aspects and path integral techniques,''}
	Trieste HEP Cosmol. 1993: 175-244, hep-th/9310144.}
\lref\ramoore{S. Cordes, G. Moore, S. Ramgoolam, 
	{\it ``Large N 2-D Yang-Mills
 	theory and topological string theory,''} 
	Commun. Math. Phys.185 (1997) 543-619, hep-th/9402107.} 
\lref\horava{P. Horava, {\it ``Topological rigid string theory and 
two-dimensional QCD,''}\ \ \
Nucl. Phys. B463 (1996) 238-286, hep-th/9507060.} 
\lref\dougjev{M. Douglas, {\it ``Conformal field theory techniques 
	in large N Yang-Mills theory,''} RU-93-57, Presented at 
Cargese
	Workshop on Strings, Conformal Models and Topological Field 
Theories, 
	Cargese, France, May 12-26, 1993, hep-th/9311130.}
\lref\mukai{S. Mukai, Invent. Math.7 (1984) 189. }
\lref\nahmhimself{W. Nahm, {\it ``The construction of all self-dual 
	multimonopoles by the ADHM method,'' } 
	Phys. Lett.90B (1980) 413.}	
\lref\morepoly{A. Polyakov, {\it ``String theory and quark 
confinement,''}
		hep-th/9711002. 
		{\it ``Confining strings,''} Nucl.Phys.B486 
		(1997) 23-33, hep-th/9607049. }
\lref\rusakov{B. Rusakov, {\it ``Loop averages and partition functions
		in U(N) gauge theory on two-dimensional manifolds,''}
		Mod.Phys.Lett.A5 (1990) 693-703.}
\lref\gtr{O. Ganor, W. Taylor, and S. Ramgoolam, {\it ``Branes,
		fluxes and duality in M(atrix) theory,''}
		Nucl.Phys.B492 (1997) 191-204, hep-th/9611202.} 
\lref\thooftor{G. 't Hooft, {\it ``Some twisted selfdual solutions 
for the
	Yang-Mills equations on a hypertorus,''} 
	Commun.Math.Phys.81 (1981) 267-275.} 
\lref\decon{B. Svetitsky and L. Yaffe, {\it ``Critical behavior at
	finite temperature confinement transitions,''} 
	Nucl.Phys.B210 (1982) 423.}
\lref\maldads{J. Maldacena, {\it ``The large N limit of superconformal 
field theories and supergravity,''} e-Print Archive: hep-th/9711200.} 
\lref\adskleb{S. Gubser, I. Klebanov, and A. Polyakov, {\it ``Gauge theory
correlators from noncritical string theory,''} 
e-Print Archive: hep-th/9802109.} 
\lref\ads{E. Witten, {\it ``Anti-de-Sitter space, thermal phase
transition,  and confinement in gauge theories,''}
e-Print Archive: hep-th/9803131. } 
\lref\mqcd{E. Witten, {\it ``Branes and the dynamics of QCD,''}
Nucl.Phys.B507 (1997) 658-690, 
e-Print Archive: hep-th/9706109.}
\lref\dougkazak{M. Douglas and V. Kazakov, {\it ``Large N phase 
transition in continuum QCD in two-dimensions,''} 
Phys. Lett.B319 (1993) 219-230,
e-Print Archive: hep-th/9305047} 
    

\newsec{ Introduction } 

The purpose of this paper is to present evidence suggesting 
the existence of a $GL(2,Z)$ duality of confining large $N$ 
gauge theory which resembles T-duality of a string description. 
We will consider $SU(N)/Z_N$ gauge theory with
two spatial directions compactified on a torus.  For simplicity
we take this torus to be square.
This theory has superselection sectors labled by the t'Hooft
magnetic flux $m$ through the torus \thooft.
We shall study the t'Hooft large $N$ limit with $g^2N$ and 
$m/N$ fixed,  and focus almost entirely on the planar limit.  
Supersymmetry will 
play no role in our present discussion.  Unlike Olive Montonen 
duality,
too much supersymmetry ruins the conjecture,  which depends
critically on confinement.  
  
Two generators of the conjectured $GL(2,Z)$ are trivially realized
even at finite $N$.  One such generator takes $m$ to $m+N$.  This
is a symmetry because the  
t'Hooft twist $m$ is only defined modulo $N$.  The other trivial 
generator
corresponds to parity and takes $m$ to $-m$.
If one could also exchange $N$ and $m$,  then  
a $GL(2,Z)$ symmetry would be generated.  
Such an exchange is often referred to
as Nahm duality.  Nahm duality \nahmhimself\nahm\ has been studied in 
the context of non-confining super Yang-Mills theories with $16$ 
supersymmetries \verlindehac.  In this case $GL(2,Z)$ can not be 
an exact duality of the theory,  in part because exchange of 
$N$ and $m$ does not make sense for $m=0$\footnote*{Yang-Mills theory 
on a non-commutative torus may have an exact duality 
analogous to Nahm duality \noncom\hulldoug.}.   
This difficulty does not 
arise in our conjecture,  for which the
modular parameter is not $\tau = {m\over N}$, but  
\eqn\mymodpar{\tau = {m\over N} + i \Lambda^2 A} 
Here $A$ is the area of the torus and $\Lambda^2$ is the string tension.
$GL(2,Z)$ is generated by
\eqn\gener{\eqalign{ \tau \rightarrow \tau + 1,\cr
		\tau\rightarrow -\bar{\tau},\cr
		\tau\rightarrow -{1\over \tau}}}
Since $m$ and $N$ are integers, the last generator makes
sense for non-zero $\Lambda^2A$ only if $N$ is taken to infinity.
In this limit $m \over N$ becomes a continuous parameter.  

In section 2 we will attempt to motivate
the conjecture by discussing qualitative  properties of 
confining Yang-Mills theory which 
are consistent with T-duality of a string description. 
The $GL(2,Z)$ we propose resembles string T-duality if one 
identifies $m\over N$ with a two form modulus in the string 
description.

If this duality exists for pure QCD,  then large $N$ pure 
QCD on $R^4$ is dual 
to a large N QCD on $R^2$ with two adjoint scalars.  
Some evidence for this is discussed in section 3.
At the same time we mention some possible problems which 
could invalidate $GL(2,Z)$ duality.
It may be that T-duality relates theories in the same 
universality class as QCD,  but is not a self duality of 
conventional QCD.  

In section 4
we discuss the status of the conjecture for pure two dimensional QCD
on $T^2$,  for which the large $N$ partition function is calculable.  
In this case the partition function is indeed a function of the 
modular parameter $\tau = {m\over N} + i{\lambda\over 2\pi} A$,
with $\lambda = g^2N$.
The partition function is almost but not exactly modular invariant.
A very simple modification of the partition function removes the 
anomaly. 

There are domains in which the proposed $GL(2,Z)$ takes weak coupling 
to weak coupling,
and should therefore be visible perturbatively.
This includes the limit in which
$\Lambda^2 A << m^2/N^2$. 
In section 4 we construct an 
explicit map
which relates theories with 
the same greatest common divisor of 
$N$ and $m$ in the limit of vanishing coupling.  
This map treats $m\over N$  as a modular parameter.
Under $\tau\rightarrow -{1\over\tau}$ the area is mapped linearly,
instead of being inverted.  In this respect
our proposal differs from Nahm duality,  which  simultaneously 
exchanges rank with flux and inverts the area.

\newsec{Duality and confinement}

It has been suspected for some time that confining Yang-Mills 
theories have
a string description \polyakov.   Pure two dimensional
QCD is known to be a string theory \gross\gtaylor\ramoore\horava,
except on a sufficiently small 2-sphere \dougkazak.  
In higher 
dimensions the situation is less clear,  although some progress has 
been
made \morepoly\maldads\adskleb\ads.  
In this section we will assume that a string 
description 
exists,  and discuss some qualitative features of a confining large N 
Yang-Mills theory on $T^2 \times R^n$ which suggest that this 
string theory may have a T-duality.  For a review of T-duality in
the context of critical strings,  the reader is referred to \tdlty.

The T-duality group for
compactifications on $T^2$ includes a $GL(2,Z)$ subgroup for which 
the 
modular parameter is $\tau = B + i \Lambda^2 A$.  $B$ is a flat two 
form background
on $T^2$, $A$ is the area of $T^2$, and $\Lambda$ is the string 
tension.
It is convenient to use complex coordinates and define 
$w = w^1 + i w^2$ where $w^i$ are the string winding numbers, 
and $P = P_1 + i P_2$ where $P_i$ are the integer string momenta.
The generator $\tau\rightarrow -{1\over\tau}$ is accompanied by
the exchange $w\leftrightarrow P$. Parity takes 
$\tau\rightarrow -\bar\tau$,
$w\rightarrow\bar{w}$ and $P\rightarrow\bar{P}$.
The remaining generator, 
$\tau\rightarrow\tau +1$, is accompanied by $P\rightarrow P + iw$. 
This momentum shift arises because the canonical string momentum $P$ 
has a 
contribution $iBw$ arising from the two form term in the world sheet 
action,
$2\pi \int_\Sigma B$. The $GL(2,Z)$ invariant energy 
of a multiplet
of mass $M$ is given by 
\eqn\energy{E = \sqrt{M^2 + {\Lambda^2\over Im(\tau)}|P-i\tau w|^2}.}

Now consider an electrically confining Yang-Mills theory
on $T^2 \times R^n$  with the time direction lying in $R^n$.
The energy of an t'Hooft electric flux \thooft\ on a large (square)
torus is given by the confining potential $e \Lambda^2\sqrt{A}$.  
This energy equals that of 
a Kaluza Klein mode on a torus with the area inverted, 
$A^{\prime} = 1/(\Lambda^4 A)$.
This is consistent with T-duality for vanishing $B$ if one 
identifies the 
electric flux with the string winding number.    
A more difficult question is whether the energy of an electric flux
on a small torus is equal to the energy of a state with momentum on
a large dual torus.  This question will be discussed in the next 
section.

One must also identify the  
quantity in the Yang-Mills theory corresponding to the two form 
modulus.
This quantity should be continuous and periodic.  A natural candidate
is $m \over N$, where $m$ is the $SU(N)$ t'Hooft magnetic flux, 
which is defined
modulo $N$. $m/N$ becomes a continuous parameter in the  
$N\rightarrow\infty$ limit.
With this identification, the two form contribution to 
the canonical string momentum 
$P_i = B_{ij}w^j +\dots$ has the form one expects
when written in terms of Yang-Mills variables;
$P_i = {m_{ij}\over N} e^j +\dots$.

This last point requires some clarification.
The division
of the momentum into a contribution from $SU(N)$ t'Hooft fluxes
and a contribution from 
everything else is more natural in if one considers instead
a $U(N)$ theory, which has vanishing total twist. 
The $U(N)$ theory is 
locally $U(1) \times SU(N)/Z_N$. 
The requirement of vanishing total 
twist amounts to the 
statement that the $U(1)$ magnetic flux is equal,
modulo N, to the $SU(N)$ magnetic fluxes \zbranes;
\eqn\mod{m^{\prime} = m + lN.} 
A similar statement,
\eqn\alsoe{e^{\prime} = e +kN,}
holds for the electric fluxes,
defined as eigenvalues
of certain large gauge transformations \thooft.
The Yang-Mills momentum is 
\eqn\momt{\eqalign{P^{YM}_i = \int Tr F_{ij}F^{0j} =\cr
	           p_i + {m^{\prime}_{ij}\over N}{e^{\prime}}^j}}
where the second term is the zero-mode contribution, which comes from
the $U(1)$ electric and magnetic flux quantum numbers.
The $U(1)$ fluxes $e^{\prime},m^{\prime}$ are well defined, however 
the $SU(N)$ t'Hooft fluxes $e$ and $m$
are defined modulo $N$. 
Therefore the momentum defined by 
\eqn\strim{P_i = p_i + {m_{ij}\over N}e^j}
has the same transformation properties under integer shifts of
$m\over N$ as the canonical string momentum.
We are not really interested in the  $U(N)$ theory however,  which in
dimensions greater than two has 
a decoupled massless photon having nothing to do with the QCD string.  
In the $SU(N)/Z_N$ case, it is less clear how to define a momentum 
with the correct transformation properties,  since the fluxes are 
not related to integrals of local operators and do not contribute
to the momentum in any direct way.  
One possibility is to define the quantity $p_i$ 
as $p_i = P_i^{YM} - {m_{ij}\over N}e^j$,  
for $-N/2 < m < N/2$.  Then
the momentum \strim\ has the correct transformation properties 
under $\tau\rightarrow\tau+1$ and parity.

Note that there is another  context in which 
momentum and electric flux are
exchanged by a T-duality \taylor\gtr.  This arises for compactifications of 
M theory on $S_1^+\times  T^d$,  which is believed to be described by a 
{\it non-confining}
$d+1$ dimensional $U(N)$ Yang-Mills theory with maximal
supersymmetry \bfss\lcone\taylor\gtr.
However in this case the $M$ theory spectrum does not correspond
to confinement in the Yang-Mills theory.  The 
momentum which gets exchanged with electric flux
is an $M$ theory momentum and not a momentum 
in the world volume of
the Yang-Mills theory.  Furthermore the Yang-Mills
energy corresponds to the M theory light cone Hamiltonian $P_M^+$ 
rather than $P_M^0$. 
The energy of a BPS electric flux is
equal to that of
a classical $U(1)$ electric flux spread evenly over the torus. 
On the other hand for a {\it confining} Yang-Mills
theory,  one expects the Yang-Mills energy to equal the 
string theory energy. 

\newsec{$D=4 \leftrightarrow D=2$ duality}

If the conjectured duality exists,  there is
a remarkable consequence for electrically confining large N Yang-Mills
theories
on $R^d$.  One can obtain $R^d$ from
$R^{d-2} \times T^2$ by making the torus very large.  
Under inversion of the area of the torus (for $m = 0$),
one obtains a dimensionally reduced 
theory. Thus duality would imply that  
large $N$ pure QCD in four dimensions is equivalent to
Large $N$ QCD with
two adjoint scalars in two dimensions. 
In fact such a two dimensional model
has been used to approximate the dynamics of pure QCD in $4$ 
dimensions  
QCD \daleykleb\igor\antonuccio.  The adjoint scalars in this model 
play the role of
transversely polarized gluons.    
In \antonuccio\ the spectrum of this
two dimensional model,  computed by discrete light cone quantization,
was compared to the glueball spectrum of pure 4-d QCD computed using
Monte-Carlo simulation.  The degree of numerical accuracy allows only
crude comparison,  however the spectra have some qualitatively 
agreement.
Perhaps in the $N\rightarrow\infty$ limit the agreement is more than 
just qualitative.

If such an equivalence exists,  duality must map the 
QCD scale nontrivially.  The mass gap is only proportional to the
QCD scale,  defined in terms of the running coupling,  if
the torus is very large.  In general
the mass gap, or string tension $\Lambda^2$,  depends on both the QCD
scale $\Lambda_{QCD4}$ and the area of the torus.
Let us fix the mass gap and take the size of the torus to infinity.
$\Lambda_{QCD4}$ on the small dual torus with area $A$ can be 
found by  
matching the coupling of the two dimensional 
reduced theory to the 4 dimensional
running coupling at the Kaluza Klein scale;
\eqn\match{\sqrt{A} g_{2D} = \sqrt{A}\Lambda = 
g_{4D}({1\over\sqrt{A}}, \Lambda_{QCD4}).}

The notion that large $N$ can generate extra dimensions is not novel.
The $d=2 \leftrightarrow d=4$ equivalence we suggest here is similar
to Eguchi-Kawai reduction \eguchi,  although our discussion pertains
to the continuum theory.
Assuming the proposed duality exists, the momentum in the 
two hidden dimensions of the reduced theory must correspond to the
electric flux
on the vanishingly small torus in the unreduced theory.
More precisely,  the hidden momentum should be related to the 
length of a string wrapped around the small torus,
$p_i = e_i \sqrt{A} \Lambda^2$,  which is held fixed as 
$A\rightarrow 0$.
For $SU(N)/Z_N$,      
the flux $e_i$ is defined \thooft\ 
by considering certain large gauge transformations $\hat{T_i}$ 
which leave 
the boundary
conditions on the torus invariant. 
Such gauge transformations correspond to elements of $SU(N)$ 
satisfying
\eqn\large{g_i(x+2\pi R_s) = g_i(x)e^{2\pi i \over N},}
where $R_s$ is the  radius of a cycle of the small (square) torus.
$\hat{T_i}^N$ is a ``small'' gauge transformation leaving physical 
states invariant.  The electric fluxes are defined by the  
eigenvalues of $\hat{T_i}$,  $\exp (2 \pi i {e_i\over N})$.   
To see what 
$e_i$ becomes in the 
reduced theory,  consider a Wilson loop around the i'th cycle 
of $T^2$.
Under a large gauge transformation
\eqn\trans{ P e^{i \oint A} \rightarrow 
	e^{2\pi i\over N}P e^{i\oint A},}
In the reduced theory, this transformation becomes a
a global symmetry:
\eqn\redtrans{ e^{i R_s X^i} \rightarrow  
	e^{2\pi i\over N}e^{i R_s X^i}.}  
Here $X^i$ are the adjoint scalars of the reduced theory.
If hidden momenta exist in the $d=2$ theory, they should be the 
generators of this 
transformation.  Since $N$ such transformations 
give the identity
map,  we wish ultimately to interpret them as translations around
a large discretized hidden torus which becomes continuous as 
$N\rightarrow \infty$. 

A problem arises because the naive reduction of pure $SU(N)$ gauge theory,
\eqn\naive{S = {N\over\lambda_{2d}}\int d^2 x Tr( F_{\mu\nu}^2 + 
	D_{\mu}X^iD^{\mu}X^i + [X^2,X^3]^2),}
is not a finite theory and requires a mass counterterm.  It also does
not have a symmetry corresponding to \redtrans.  It is 
tempting to consider
a $U(N)$ theory instead,  in which case 
the naive reduced theory has a continuous $U(1) \times U(1)$ symmetry.
However in this case there is a free photon and no mass gap.  
States charged under
the  $U(1) \times U(1)$  have energies
inconsistent with \energy.  Therefore we will only consider 
the $SU(N)$ theory. 
For $R_s$ finite but small compared to $1\over\Lambda$,
an effective action which is $Z_N \times Z_N$ symmetric 
may be written in terms of the $SU(N)$ Wilson loops 
$h_i = \exp(iR_sX_i)$;
\eqn\goodaction{S_{SU(N)} =  
	{N\over\lambda_{2d}}\int d^2 x Tr \left( F_{\mu\nu}^2 
	+ {1\over R_s^2} (h_i D_{\mu} h_i^{\dagger})^2 + 
	{1\over R_s^4} [h_2,h_3][h_2^{\dagger},h_3^{\dagger}]\right).}
The naive reduced action is recovered by writing 
$h_i = 1 + \exp(iR_sX^i) +\dots$ and taking $R_s\rightarrow 0$ with
$X_i$ fixed. The metric of this sigma model is proportional to 
$1/ R_s^2$,  which corresponds to the area of the hidden 
torus on which
$Z_N \times Z_N$ acts as a translation. 
Note that a mass counterterm proportional to 
${1\over R_s^2} \sum_i Tr h_i$
is prohibited by the $Z_N\times Z_N$ symmetry.   

The important question is whether this 
symmetry is spontaneously broken below a critical $R_s$.  
The
two dimensional theory can not generate extra dimensions
as $N\rightarrow \infty$ if $R_s < R_s^c$.
If there is a T-duality,  the dual torus should have an unbroken 
translation symmetry.
However, from experience with finite temperature deconfinement 
transitions,  
one might conclude that $Z_N \times Z_N$ should be 
broken for sufficiently 
small $R_s$.  
Note that in our case, symmetry breaking would 
not be interpreted
as deconfinement,  since the order parameters are spacelike Wilson
loops,  rather than a timelike Wilson loop\footnote*{In 
the finite temperature theory,  the Wilson loop wrapped
around the Euclidean time direction is an order parameter 
of the deconfining 
phase transition \decon.  In this case,  the $Z_N$ symmetry is 
unbroken
in the confining phase. }. 

The question of whether symmetry breaking occurs at finite $R_s$
is closely tied to the question of whether the limit we wish to take
exists.  This limit is $N\rightarrow\infty$ followed by
$R_s\rightarrow 0$ while tuning $\Lambda_{QCD4}$ to keep the 
mass gap fixed.  
If $Z_N$  can be thought of as a
continuous $U(1)$ in the $N\rightarrow\infty$ limit, 
then the broken phase would have a goldstone boson.
Thus there can not be any symmetry breaking if the reduced theory 
has a gap in the large $N$ limit.
Furthermore if a string description remains valid for small $R_s$,
one would not expect symmetry breaking for $N\rightarrow\infty$.
If there were symmetry breaking, states with different
$Z_N\times Z_N$ charges, or  ``hidden momenta,'' 
would become degenerate and condense.   
Recall that the hidden momenta are given by $e_iR_s \Lambda^2$.
$e_iR_s$ is the minimum length of a QCD string wrapping $e_i$ times 
around a cycle of the small torus.
Since $e_i$ is defined modulo $N$, the minimum length can be at most
$(N-1)R_s$.  If $N$ is infinite,  then the minimum length is unbounded
and can be held fixed as $R_s \rightarrow 0$ by scaling $e_i$ like 
$1/ R_s$.  For fixed string tension $\Lambda^2$ one would expect  
wrapped strings with arbitrarily large minimum lengths to be heavy,  
in which case they could not condense.  


If there is no symmetry breaking at infinite $N$,  then 
the vacuum 
has zero electric flux.  Introducing even 
the minimum amount of flux, or a singly wrapped string,  
would lead to an energy in excess of the mass gap. 
A possible explanation for such unusual behavior
is that the string spreads in the non-compact direction as 
$R_s\rightarrow 0$. 

If it can be shown that the $Z_N \times Z_N$ symmetry is unbroken in 
the large $N$ limit,  one must then show that the spectrum of the 
$R_s\rightarrow 0$ theory 
has a hidden four dimensional Lorentz invariance.  
This might be testable numerically using the two dimensional
description.  Having 
found candidates for momentum in the hidden directions,
it might also be possible to see if a four dimensional Lorentz algebra 
exists. However we leave these tests for the future.

It is entirely possible that the $Z_N \times Z_N$ symmetry is
spontaneously broken in conventional pure QCD on a sufficiently 
small torus.  Nonetheless,  $GL(2,Z)$ might still relate theories
in the same universality class as QCD. 
Also if the reduced theory \goodaction,  has a symmetry 
breaking transition at finite  $R_s$,  it could be interesting 
to study this theory for $R_s > R_s^c$,  since in this domain 
the large $N$ limit may generate extra dimensions.  
This action is also interesting since it appears to be 
a more suitable action to describe dimensionally reduced pure QCD,
due to the absence of a mass renormalization.

\newsec{Modular invariance in two dimensions}

	While the QCD string may fail to have a T-duality in four 
dimensions,  it comes very close to having a T-duality in two dimensions.
In this section we study pure Euclidean $SU(N)$ Yang-Mills theory on
$T^2$ in the t'Hooft large $N$ limit.  Although this theory has no
dynamics,  we can use it to test the conjecture that the t'Hooft twist
and the area combine into a modular parameter \mymodpar.
The partition function of this 
theory on a surface of arbitrary genus is known in terms of a 
sum over 
representations of $SU(N)$ \rusakov\migdal. 
On a two torus,   the partition
function in the absence of twisted boundary conditions is given 
by 
\eqn\twod{Z = \sum_R e^{g^2AC_2(R)},}
where $C_2(R)$ is the quadratic casimir in the representation $R$.
In a sector with t'Hooft twist
$m$, the partition function may be computed by continuum 
methods \witten,  and is also easily computed using the heat
kernel action on a $T^2$ lattice with twisted boundary conditions.  
The result is
\eqn\twodtwist{Z = \sum_R e^{g^2AC_2(R)} {{\cal{X}}_R(D_m)\over d_R}}
where ${\cal{X}}_R (D_m)$ is the trace of the 
element $D_m$ in the center of $SU(N)$ corresponding to the t'Hooft 
twist.
In a representation whose Young Tableaux has $n_R$ boxes,  
\eqn\twistrep{D_m = e^{2 \pi i {m\over N} n_R}}
To compute the large $N$ expansion of the partition function,  
we repeat
the calculations of \gtaylor for the case of nonvanishing twist.
This expansion is obtained by considering composite
representations $\bar{S}R$ obtained by gluing the Young Tableaux of a 
representation $R$ with a finite number of boxes onto the 
right of the 
complex conjugate of a representation $S$ with a finite 
number of boxes.
The quadratic Casimir of such a representation in the ${1\over N}$ 
expansion is 
\eqn\casimir{C_2(\bar{S}R) = n_RN + n_SN + \dots}
Furthermore,
\eqn\dm{D_m = e^{2 \pi i {m\over N} (n_R - n_S)}}
Therefore the free energy at leading order is 
\eqn\leading{F = ln \left|\sum_n^{\infty} 
	\rho (n) e^{g^2NAn + 2 \pi i {m\over N} n}\right|^2}
where $\rho (n)$ counts the number of representations with $n$ boxes.
This sum is computed
just as in \gross,  giving
\eqn\zmodular{F = ln  \left| {e^{2 \pi i {\tau\over 24}} 
		\over \eta(\tau)} \right| ^2}
where $\eta$ is a Dedekind eta function,  and 
\eqn\modparam{\tau = {m\over N} - {\lambda A \over 2\pi i},}
with $g^2N = \lambda$.   Thus the complexification of the 
area generated 
by modular 
transformations corresponds to a non-zero t'Hooft twist.
Upon completing this paper we became aware that 
M. Douglas has also made this observation using a Jevicki-Sakita
boson description of 2-d QCD on the torus \dougjev.
As noted in \rudd\dougjev\  
the free energy is almost,  but not exactly 
invariant under inversion of the area. 
The eta function has the modular properties
\eqn\etaprop{\eqalign {\eta (\tau + 1) = \eta (\tau)\cr
			\eta(-{1\over\tau}) = 
\sqrt{i\tau} \eta(\tau).}}
A simple modification of the partition function,
\eqn\trumod{ {\cal{Z}} = Z {e^{\lambda A\over 24}\over 
\sqrt{\lambda A}}}
is modular invariant. 
The extra factor exponential in the area is a local
term,
$\sim\int\sqrt{\det g}$,  corresponding to a modified ground state 
energy. 
On the other hand, the non-local $\sqrt{\lambda A}$ factor poses 
a problem for modular 
invariance.  Nonetheless the 
deviation from
modular invariance is very simple and we feel deserves better 
understanding.  
The two dimensional QCD string has at least an ``approximate'' 
T-duality.

\newsec{Exchange of rank and twist}

The $GL(2,Z)$ we have proposed treats $m\over N$ as a modular 
parameter
as long as one only acts with $\tau\rightarrow -1/\tau$ in the region
$\Lambda^2 A N^2 / m^2 << 1$.
In this domain the area is mapped linearly and is never inverted.
If this duality exists,  it should be possible to see it 
perturbatively
in this region.    
In this section we discuss an attempt to construct 
a classical map under which $N$ and $m$ transform as a doublet of 
$GL(2,Z)$.
We are only able to show the validity of the map we construct 
in the limit of vanishing coupling.  However if it does extend 
to finite coupling,  
it has the correct qualitative property that the area of the torus 
is mapped linearly rather than being inverted. 
While the results of
this section are very far from quantitative rigor,  we feel they are 
at least suggestive. 
The map we propose will take an $SU(N)/Z_N$ theory with flux $m$,  
or an $(N,m)$
theory,  into a $(p,0)$ theory where $p$ is the greatest common 
divisor of
$N$ and $m$.  $GL(2,Z)$ is then generated by inverting this map 
to get 
other theories with the same greatest common divisor. 

We begin by reviewing  the 
definition of magnetic flux for $SU(N)/Z_N$ Yang-Mills on a torus.
Following t'Hooft \thooft, translations around a cycle of the torus 
are 
equivalent to gauge transformations:
\eqn\tw{A_i(x+a_j) =  U_j(x)A_i(x)U_j^{\dagger}(x)
			+ U_j(x)i\del_i U_j^\dagger(x)}
where $a_i$ are the periodicities of the torus.  
For adjoint matter,  one has the following constraint: 
\eqn\part{U_i (x) U_j (x+a_i)  
		U^{\dagger}_i (x+a_j)
		U^{\dagger}_j (x) = e^{2\pi i {m_{ij}\over N}}}
where the twist $m_{ij}$ is an integer,  and is taken as the 
definition of 
nonabelian 
magnetic flux.  We consider the case of Yang-Mills on 
$T^2 \times R^n$, and drop the indices $ij$.
Note that $m\rightarrow m+jN$ is a 
manifest symmetry.  We shall assume the timelike 
direction lies in $R^n$.  
In $A_0 = 0$ gauge we choose
the following twisted boundary conditions 
\eqn\chose{ \eqalign{ U_1(x) = Q, \cr
		      U_2(x) = P^m,}}
where 
\eqn\wepick{
Q = e^{i\theta}
\pmatrix{
1& & & \cr
& e^{2\pi i\over N} & & \cr
& & \ddots & \cr
& & & e^{2\pi i (N-1)\over N}
},\qquad
P = e^{i\theta^{\prime}}
\pmatrix{
 & 1 & & \cr
 &   & 1 & \cr
 &   &   & \ddots \cr
1&   &   &  
}}
The phases $\theta$ and $\theta^{\prime}$ are chosen so that
$Q$ and $P$ have determinant $1$.
$Q$ and $P$ satisfy 
\eqn\QPQP{PQ = QP e^{2\pi i\over N}.}
Different choices of twisted boundary conditions with the same 
magnetic flux
are related by large gauge transformations,  but in $A_0=0$ gauge no
paths connect different twisted boundary conditions at different 
times.
Different twisted boundary conditions with the same 
magnetic flux
define equivalent superselection sectors and there is nothing 
special about
our choice.  There are other large gauge 
transformations which leave the boundary conditions invariant.  The 
eigenvalues of
of these gauge transformation determine the t'Hooft electric flux.

The constraints imposed by twisted boundary conditions can be 
solved to 
find the independent perturbative degrees of
freedom.  To this end we shall work in momentum space.  Since 
$Q^N = 1$,  
the gauge potentials are  
periodic in $x^1$ on the interval $[0,Na^1]$.  To find the
periodicity in $x^2$,  one looks for the minimal power to 
which one must
raise $P^m$ to get $1$.  Writing the pair $(N,m)$ as 
$(p\alpha,p\beta)$
where $\alpha$ and $\beta$ 
are relatively prime,  one finds that this power is $\alpha$. 
Therefore the 
gauge potentials are periodic
in $x^2$ on the interval $[0, \alpha a_2]$.  The fourier 
modes
of the gauge field strength $F_{n_1,n_2}$ satisfy the twisted 
boundary 
conditions, 
\eqn\twstone{e^{2\pi i {n_1\over N}} F_{n_1,n_2} = Q 
F_{n_1,n_2} Q^{\dagger}}
and
\eqn\twsttwo{e^{2\pi i {n_2\over \alpha}} F_{n_1,n_2} = 
		P^m F_{n_1,n_2} P^{m\dagger}.}
Making use of the algebra \QPQP, a general solution of \twstone\ is
\eqn\solone{ F_{n_1,n_2} = M_{n_1,n_2} P^{n_1},}  
where $M_{n_1,n_2}$ is a diagonal $N \times N$ matrix which is 
traceless
when $n_1 = 0$ mod $N$.  
Note that if $m$ vanished  there would be $N-1$ degrees of freedom
for each fourier mode with $n_1=0 mod N$ and $N$ degrees of freedom 
for all the others,  rather than 
$N^2-1$ degrees of freedom for every fourier mode.  
This is because the (non-gauge invariant) momentum
in the $x^1$ direction of the torus is fractional in units of 
${1 \over N}$.  Heuristically,  in going to a more conventional 
gauge with 
$U_1 = U_2 = I$,  the fractional modes become integral and
fill out the Lie algebra.  

Now consider arbitrary $m$.  The second constraint \twsttwo\
gives
\eqn\mconst{ P^m M_{n_1,n_2} 
	P^{-m} = M_{n_1,n_2} e^{2\pi i {n_2\over \alpha}},}
Conjugating $M$ by $P^m$ shifts the elements of $M$ cyclically
by an amount $m$.   Thus we find that the number of independent 
elements
of $M_{n_1,n_2}$ is $p$, the greatest common divisor of $N$ and $m$; 
\eqn\dualf{M_n = 
\pmatrix{M_n^{\prime} &  &  & \cr
	 & M_n^{\prime}e^{2\pi i {n_2 \over \alpha \beta}} & & \cr
	 & & \dots & \cr
	 & & & &\cr}}
where $M_n^{\prime}$ is a diagonal $p \times p$ matrix. 
It is now easy to construct a candidate for 
the gauge field of the 
dual theory with rank $p$ and vanishing magnetic flux, or
a $(p,0)$ theory.
We define the dual field strength as 
\eqn\thedual{F^{\prime}_n = e^{i\phi(n)}M^{\prime}_n P^{\prime n_1}}
where $P^{\prime}$ is the $p \times p$ shift matrix.  The diagonal 
phase factor 
$\phi$ is chosen so that the dual field strength is real, 
$F^{\prime}_{-n} = F^{\prime\dagger}_{n}$.
This is general solution  of the constraint
\eqn\twstoneprime{ e^{2\pi i {n_1\over p}} F^{\prime}_n = 
	Q^{\prime} F^{\prime}_{n_1,n_2} Q^{\prime\dagger},}
where $Q^{\prime}$ is the $p \times p$ matrix
\eqn\qprime{      
 Q^{\prime} = 
\pmatrix{
1& & & \cr
& e^{2\pi i\over p} & & \cr
& & \ddots & \cr
& & & e^{2\pi i (p-1)\over p}
}.}
Therefore $F^{\prime}$ is a candidate for a field strength on a 
dual torus 
with the twisted boundary
conditions given by $U^{\prime}_1 = Q^{\prime}$ and 
$U^{\prime}_2 = I$, 
which corresponds to vanishing magnetic flux.    

The action of the $(N,m)$ theory is
\eqn\action{S = \int_{R^n} \sum_n Tr_{N\times N} F_{\mu\nu, n} 
				F^{\dagger}_{\mu\nu, n} 
		} 
Written in terms of the proposed dual field strength this becomes
\eqn\dualaction{S = \int_{R_n} \alpha Tr_{p \times p} 
F^{\prime}_{\mu\nu, n}
				F^{\prime\dagger}_{\mu\nu, n}.}
However,  for duality to hold,  we must be able to define a dual 
gauge potential which solves the Jacobi identity and gives the 
correct
measure in the path integral.  Let us define the
The dual gauge potential the same way we defined the the dual field 
strength.  Then the map between the $(N.m)$ gauge potential and 
the dual
$(p,0)$ gauge potential is linear, so one might expect
that the measure maps correctly.  
We will scale the fields such that the coupling constant appears only
in the interaction terms.
If one neglects the
the $[A_{\mu},F^{\alpha\beta}]$ terms, it is easy to check 
that the Jacobi identity is preserved by the map,  provided that that
the periodicities of fields on the original torus are the same as 
those 
on the dual torus.
In other words $Na^1 = N^{\prime}{a^{\prime}}^1 = 
p{a^{\prime}}^1$ and 
$\alpha a^2 = \alpha^{\prime}{a^{\prime}}^2 = {a^{\prime}}^2$.
Thus the area of the torus is mapped linearly.
Working in a Hamiltonian
formulation, one can easily check that the Hamiltonian,  
commutation relations, and
Gauss law constraint of the $(N.m)$ theory at zero coupling 
map to those of the $(p,0)$ theory at zero coupling. 

At finite coupling however,  the Jacobi identity is no 
longer satisfied.  It
is violated by terms involving the difference between the 
$N \times N$ shift
matrix $P$ and $\alpha$ copies of the $p \times p$ shift matrix, 
\eqn\differ{ P - \pmatrix{ P^{\prime} & & \cr
			   &  P^{\prime} & \cr
			   &  &  \dots}.}
These matrices are related by moving
a finite number of ones and zeros in the limit that 
$p \rightarrow \infty$.
Thus it is very tempting to 
neglect the difference. 
However when these matrices are raised
to a power of order $p$,  the difference is not always negligible.  
This occurs when $n_1$ is of order $p$.  We can not discard
such modes from the action,  since they may correspond to a 
finite gauge 
invariant physical momenta.  
It may be possible that this discrepancy is negligible at 
leading order in some
weak coupling expansion,  however we will not attempt to prove this.

\newsec{Conclusion}

We have found properties of large N confining Yang-Mills theories 
which are suggestive of a duality resembling T-duality of a string
description. Whether such a duality really exists for pure QCD or 
some theories in the same universality class remains to be seen.
If such a duality existed it would be quite useful,  since it would
relate theories on $R^4$ to more numerically tractable theories on
$R^2$. Among the the principle obstacles to such a duality is 
the possibility of $Z_N \times Z_N$ symmetry breaking,  and the 
absence of a string description on a sufficiently small torus.  
If the $Z_N \times Z_N$ symmetry is broken at finite radius,
there may still be interesting phenomena at large $N$ for $R>R_c$
due to the existence of wrapped string states.
Also,  the two dimensional gauged sigma model given by \goodaction
may have interesting properties for $R_s > R_s^c$,  such as the 
generation of extra dimensions in the large $N$ limit.

\bigbreak\bigskip\bigskip
\centerline{\bf Acknowledgments}\nobreak

I am especially grateful to J. Nunes, S. Ramgoolam, and W. Taylor for
very helpful discussions.  I am also indebted to  C. Callan,  
A, Cohen, 
A. Jevicki,  I. Klebanov,  and M. Schmaltz. 
This work was supported in part by NSF grant PHY96-00258.

\listrefs

\end